\documentclass[acmsmall,nonacm]{acmart}
\AtBeginDocument{%
  }

\usepackage{braket}
\usepackage{siunitx}
\usepackage{makecell}
\usepackage{subfigure}
\usepackage{subcaption}
\usepackage{graphicx}

\begin{document}

\title{Simulation of Quantum Computers: Review and Acceleration Opportunities}

\author{Alessio Cicero}
\email{alessio.cicero@chalmers.se}
\affiliation{%
  \institution{Chalmers University of Technology and University of Gothenburg}
  \city{Gothenburg}
  \country{Sweden}
}

\author{Mohammad Ali Maleki}
\email{mohammad.ali.maleki@chalmers.se}
\affiliation{%
  \institution{Chalmers University of Technology and University of Gothenburg}
  \city{Gothenburg}
  \country{Sweden}
}

\author{Muhammad Waqar Azhar}
\email{waqar.azhar@zptcorp.com}
\affiliation{%
  \institution{ZeroPoint Technologies AB}
  \city{Gothenburg}
  \country{Sweden}
}

\author{Anton Frisk Kockum}
\email{anton.frisk.kockum@chalmers.se}
\affiliation{%
  \institution{Chalmers University of Technology}
  \city{Gothenburg}
  \country{Sweden}
}

\author{Pedro Trancoso}
\email{ppedro@chalmers.se}
\affiliation{%
  \institution{Chalmers University of Technology and University of Gothenburg}
  \city{Gothenburg}
  \country{Sweden}
}

% \affiliation{%

% \affiliation{%

% \affiliation{%

% \affiliation{%

% \affiliation{%

% \affiliation{%

% \affiliation{%

\renewcommand{\shortauthors}{A. Cicero et al.}

\begin{abstract}
Quantum computing has the potential to revolutionize multiple fields by solving complex problems that can not be solved in reasonable time with current classical computers. Nevertheless, the development of quantum computers is still in its early stages and the available systems have still very limited resources. As such, currently, the most practical way to develop and test quantum algorithms is to use classical simulators of quantum computers. In addition, the development of new quantum computers and their components also depends on simulations.

Given the characteristics of a quantum computer, their simulation is a very demanding application in terms of both computation and memory. As such, simulations do not scale well in current classical systems. Thus different optimization and approximation techniques need to be applied at different levels.

This review provides an overview of the components of a quantum computer, the levels at which these components and the whole quantum computer can be simulated, and an in-depth analysis of different state-of-the-art acceleration approaches. Besides the optimizations that can be performed at the algorithmic level, this review presents the most promising hardware-aware optimizations and future directions that can be explored for improving the performance and scalability of the simulations.

\end{abstract}





\maketitle

% \section{Introduction}
\section{Introduction}
As we try to solve more and more complex problems such as developing new chemical compounds~\cite{quantum-advantage3} or evaluating the physical properties of new materials~\cite{quantum-advantage4}, the demand for computational resources continues to grow.
Certain problem cases are so complex that no existing computer system can solve them within reasonable time. For such cases, {\em Quantum Computing}~\cite{Nielsen_Chuang_2023} is a promising emerging computing paradigm that can provide solutions to these problems~\cite{superconducting-arch1, nisq-original, quantum-advantage3, quantum-advantage4, quantum-advantage5, quantum-advantage6, quantum-algorithms-review,quantum-advantage-latest}.

As quantum computing is still in its early stages of development, real machines are scarce and the existing ones have limited compute resources. Although several quantum computing experiments have shown promising results and quantum computers are being scaled up to hundreds of qubits~\cite{Arute2019, Madsen2022, Kim2023, Bluvstein2024, Acharya2024}, full quantum advantage has yet to be achieved. 

In order for algorithms to be executed in a quantum computer, they are represented as quantum circuits. The ability of quantum computers to solve increasingly complex problems is limited by the maximum size of the executable quantum circuit. 
This size is mainly bounded by two factors: (1) the number of available qubits (the fundamental unit of a quantum circuit) and (2) the circuit depth, which refers to the number of distinct timesteps at which quantum gates are applied~\cite{Nielsen_Chuang_2023}. In current quantum computer implementations, not only are qubits few but also the time they are stable (i.e., their value remains reliable) is limited. This stable time is limited by the qubit implementation technology and also their sensitivity to fluctuations in the environment (e.g., magnetic fields and temperature).

Quantum circuit execution is just one element of the {\em Quantum Computing Stack}~\cite{quantum_stack}. 
Other elements include mapping an algorithm to a quantum circuit, generating pulses to execute gates on qubits, and reliably reading out the results.
To develop and test the various parts of the quantum computing stack, accurate and fast simulation tools are essential. Similarly to classical computers, different simulation tools are needed at various design and verification steps, each with a specific focus and level of detail. 
A generic quantum computer system is composed of different parts, and the development of each part can be assisted by a different type of simulation.

This review work focuses in particular on the simulation of the execution of quantum circuits. Simulations of quantum circuit execution can be performed at different levels, from simulating the behaviour of the quantum hardware platform\cite{qtcad}\cite{Qiskit}, to simulating the interaction between a few qubits and a coupler forming a gate\cite{csqr,scqubits1,scqubits2}, and simulating entire circuits\cite{qhipster,IntelQuantumSDK,qkit, Qiskit, QuEST, HyQuas,bergholm2022pennylane}.

While several simulators have recently been deployed and made publicly available, the need to simulate more complex algorithms, circuit configurations, or more accurately model qubit behavior leads to an exponential increase in computational and memory demand for the simulations. 
Scaling up the simulations without exponentially increasing the amount of necessary resources, requires optimisations or approximations at different levels. As such, several classical computer techniques, such as data compression or optimized parallel execution,  have been applied to reduce the memory requirements and accelerate the computations~\cite{Lykov,bayraktar2023cuquantum}.

Several review works on classical simulation of quantum computers are available, but their focus is different from the acceleration of the simulation. Some works focus on the state-of-the-art numerical methods, such as the work by Xu et al.~\cite{survey-herculean} which gives a general overview, and Jones et al.~\cite{survey-distributed} which focuses on the full-state simulation techniques. Other works focus both on reporting the state of the art for  simulators as well as giving a high-level overview of the acceleration approach as in the work by Young et al.~\cite{survey-simulating}.  The work by Heng et al.~\cite{survey-exploiting} focus on presenting optimisations for GPU execution while the review by Jamadagni et al.~\cite{survey-benchmarking} focuses on setting up a benchmark for multiple simulators available, and comparing their performance. 
In contrast to the previous works, our work addresses the topic in a different way, with a goal of focusing on the possible hardware-aware acceleration techniques. It provides up-to-date details about the available simulator approaches, tools and techniques, and focuses on the possible optimisations for a broader selection of hardware platforms.

First,  this review work presents an overview of the different types of simulators for the execution of quantum circuits. The simulators are organized into different categories, to help navigate the landscape of the available tools. Second, it provides an overview of acceleration techniques described in the state of the art to improve the performance of the simulations. Having a global overview of the existing proposed solutions, we will infer what the most promising trends are and speculate on the future directions for the acceleration of the simulation of quantum computers. 
In this context, different works propose different approaches to optimize the simulation on different hardware platforms, such as Central Processing Unit~(CPU), Graphics Processing Unit~(GPU), Field Programmable Gate Array~(FPGA), or more complex setups such as a hybrid CPU and Quantum Processing Unit~(QPU) core. A summary of the hardware-aware techniques observed in the analyzed works is presented and this is used as a basis to extrapolate what the future directions in hardware-aware acceleration for quantum computer simulation should be.

The organization of this survey is as follows. Section~\ref{sec:bg} introduces the essential concepts in quantum computing necessary for understanding the rest of the sections.  In Section~\ref{sec:qc} we give an overview of the various parts of a quantum computer. Section~\ref{sec:qc-sim} covers simulations at different levels and provides an overview of the available simulators for each of them, also introducing the main bottlenecks for scaling up the number of simulated qubits. Section~\ref{sec:sim-plat} details the most common hardware platforms used for simulation. The following sections cover different acceleration methods, sorted by platform: CPU in Section~\ref{sec:cpu-acc}, GPU in Section~\ref{sec:gpu-acc}, and FPGA in Section~\ref{sec:fpga-acc}. A summary and indication of future directions for hardware-aware optimization is presented in Section~\ref{sec:summary}, and Section~\ref{sec:conclusion} concludes this work.

% \section{Background}
\section{Background}
\label{sec:bg}

This section presents a quick overview and introduction to the basic quantum computer concepts.

\subsection{Qubits}
The fundamental computational element for the quantum computer is the qubit (quantum bit). Differently from normal bits, qubits can be in a state different from just $\ket{0}$ and $\ket{1}$: they can form a linear combinations of states, usually called superposition:
\begin{equation}
\ket{\psi}=\alpha \ket{0}+\beta \ket{1} =     
    \begin{pmatrix}
    \alpha \\
    \beta 
    \end{pmatrix}
    \label{eq:qubit-equation}
\end{equation}

\begin{figure}[htbp]
    \centering
    \includegraphics[width=0.3\textwidth]{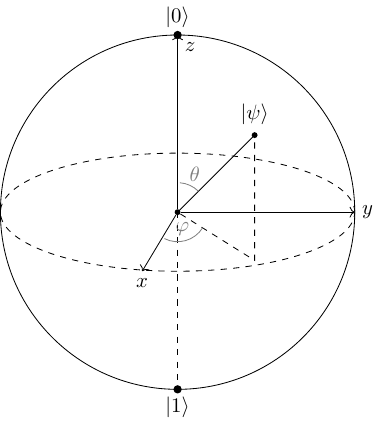}
    \caption{Bloch-sphere representation of a qubit.}
    \Description[Qubit state visualised on the Bloch sphere using polar coordinates]{The image shows the angle theta between the z-axis to the line that connects the origin of the sphere and the qubit complex value, and the angle phi between the x-axis and the line that connects the origin with the projection of the qubit complex value on the xy plane}
    \label{fig:bloch-sphere}
\end{figure}

It is possible to visualize the state of the qubit on the Bloch sphere in Figure~\ref{fig:bloch-sphere} using the equivalences~\cite{Nielsen_Chuang_2023}:
\begin{equation}
    \alpha = \cos\frac{\theta}{2}, \;\beta = e^{i\varphi}\sin\frac{\theta}{2}
    \label{eq:probability-to-angle}
\end{equation}

The vector is parametrized by two complex numbers, $\alpha$ and $\beta$, which are two probablity amplitudes. As probability amplitudes they must satisfy the property $|\alpha|^2 + |\beta|^2=1$ \cite{Nielsen_Chuang_2023}.
In order to model the value of the qubits in a classical computer, for example for simulation purposes, it is necessary to store both $\alpha$ and $\beta$ as complex values which requires a higher memory usage compared to storing the non-complex values in classical computers. 
\subsection{Quantum gates}
Quantum gates are the quantum equivalent of a logical gate. By applying a quantum gate to a qubit or multiple qubits, it is possible to control the probability amplitudes and thus change the state of the qubits.

\subsubsection{Single-qubit gates}
Operations on qubits must preserve the norm $|\alpha|^2 + |\beta|^2=1$. therefore they are described by $2 \times 2 $ unitary matrices. 
Some of the most important single-qubit gates are the Pauli gates:
\begin{equation}
    \label{eq:pauli-gates}
    X= \begin{pmatrix}
    0 && 1\\
    1 && 0
    \end{pmatrix}\; 
    Y = \begin{pmatrix}
    0 && -i\\
    i && 0
    \end{pmatrix} \;
    Z= \begin{pmatrix}
    1 && 0\\
    0 && -1
    \end{pmatrix}
\end{equation}
These matrices correspond to the rotation of $\pi$ radians around respectively the x,y, and z axes of the Bloch sphere.\par
Another important single-qubit gate is the identity matrix
\begin{equation}
\label{eq:identity-gate}
    I= \begin{pmatrix}
    1 && 0\\
    0 && 1
    \end{pmatrix}
\end{equation}
which does not affect the value of the qubit,
\subsubsection{Multi-qubit gates}
Multi-qubit gates allow $n$ qubits to interact together. The probability amplitudes required to represent an $n$-qubit system are $2^n$. For instance, a 2-qubit system can be represented as:
\begin{equation}
    \ket{\psi}= a_{00}\ket{00}+a_{01}\ket{01} + a_{10}\ket{10} +a_{11}\ket{11}
    \label{eq:multiple-qubit-state}
\end{equation}
Therefore, the matrix size for an operation on an $n$-qubit system is $2^n \times 2^n$. An important class of multi-qubit gates are the controlled gates. The control qubit values determine whether the controlled qubit's or qubits' value(s) will be controlled by the gate. An example of a  2-qubit gate in this class is the controlled-$NOT$, which applies a Pauli X gate on the target qubit if the control qubit is $\ket{1}$:
\begin{equation}
    CNOT = \begin{pmatrix}
    1 && 0 && 0 && 0 \\
    0 && 1 && 0 && 0 \\
    0 && 0 && 0 && 1 \\
    0 && 0 && 1 && 0 \\
    \end{pmatrix}
    \label{eq:cnot}
\end{equation}
Another example of multi-qubit gate is the Controlled-Z (CZ), which applies a Pauli Z gate on the controlled qubit in the case of a control qubit with state $\ket{1}$.

\subsection{Quantum circuit execution}
\begin{figure}[htbp]
    \includegraphics[]{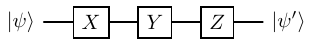}
    \caption{A single-qubit circuit, which applies the Pauli gates X, Y, and Z in succession.} 
    \Description[Quantum circuit diagram representing a quantum state psi undergoing a sequence of quantum gate operations: the X gate, Y gate, and Z gate, resulting in a new quantum state psi first.]{The image illustrates a quantum circuit where an initial quantum state psi is processed through three consecutive quantum gates: the X gate, followed by the Y gate, and finally the Z gate. Each gate is represented by a rectangular box labelled with its respective gate name (X, Y, Z). The circuit begins with the input state psi on the left and concludes with the output state psi first on the right.}
    \label{fig:one-qubit-circuit}
\end{figure}

Quantum circuits are represented as in the diagram shown in Figure~\ref{fig:one-qubit-circuit}. Quantum gates are applied to a qubit in the time domain. Thus the horizontal line connecting the different gates represent the time line (from left to right) of the different gates applied to the same qubit. In the case of a superconducting computer, this is done by controlling the single qubit with a specific pulse, and the gates are applied in successive instants of time. The circuit depth, as previously mentioned, is the longest path in the circuit, representing the maximum number of gates applied to a qubit. From a mathematical point of view, applying multiple gates can be represented as multiplying the gate matrices with the qubit state vector:
\begin{equation}
     \ket{\psi '} = Z \cdot Y \cdot X \cdot \ket{\psi} 
     \label{eq:execute-quantum-gates}
\end{equation}

\begin{figure}[htbp]
    \centerline{
        \includegraphics[]{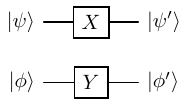} 
    }   
    \caption{A two-qubit circuit, which executes a Pauli X gate on the first qubit and a Pauli Y gate on the second qubit} 
    \Description[Two qubit circuit, similar to two single-qubit circuits stacked vertically.]{The image contains two quantum circuit diagrams, each depicting a different quantum gate operation applied to qubit states. The first diagram shows an initial quantum state psi on the left side of the X gate, represented as a rectangular box. The output of this operation is a transformed quantum state psi first on the right side of the box. The second diagram is similar, showing a quantum state labelled phi connected to the quantum gate Y also represented as a box. The output is a transformed quantum state labelled phi'.}
    \label{fig:two-qubit-circuit}
\end{figure}

In case of circuits with multiple qubits and gates, the tensor product of two gates is equivalent to executing the gates in parallel. If we consider, for example, the circuit in Figure~\ref{fig:two-qubit-circuit}, it is possible to compute the gate matrices:
\begin{equation}
    A = X \otimes Y
    \label{eq:gates-product}
\end{equation}

%The complexity of the circuit scales linearly as a function of the circuit depth, as it introduces linearly additional gates, and scales exponentially  as a function of the number of qubits, as the size of the matrix describing the gate will scale exponentially. 
The complexity of the classical model representation of a quantum circuit scales linearly with the circuit depth and exponentially with the number of qubits.
This is because with $n$ qubits there are $2^n$ possible states, and the size of the resulting matrix representing the parallel gates will be $2^n \times 2^n$. If there are $j$ successive gates, there will be the need for $j \times 2^n \times 2^n$ matrices to represent the quantum circuit.
% Circuit depth, scaling issue, memory requirements. How to apply a gate, how to apply multiple gates
% \subsubsection{Circuit simulation and emulation}

\subsection{Tensor networks}
\label{subsec:tensor-networks-bg}
Some of the simulation methods are based on the representation and execution of the quantum circuit as a tensor network.

\begin{figure}
    \subfigure[]{\label{fig:tensor-3d}\includegraphics[width=0.45\textwidth]{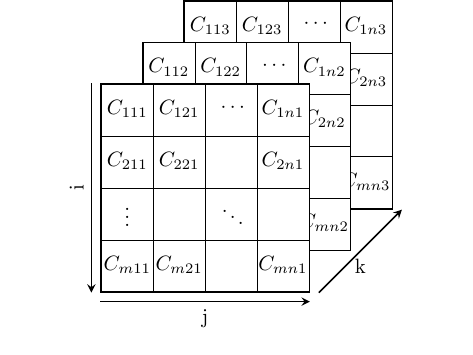}}
    \subfigure[]{\label{fig:tensor-legs}\includegraphics[width=0.45\textwidth]{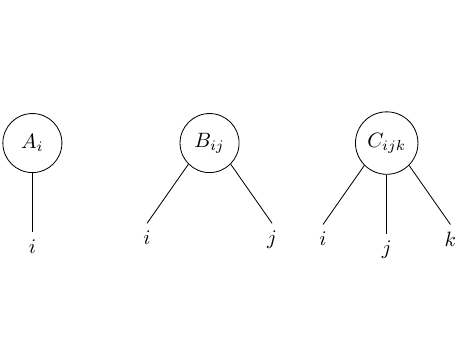}}
    \caption{(a)~Example of order-3 tensor, (b)~An order-k tensor requires k-indices to be accessed.}
    \Description[Two subfigures, on the left an order-3 tensor, shown as three matrixes, on the right the representation of the number of indices needed to access different order of tensors.]{ The left subfigure depicts a visual representation of a 3-dimensional tensor C. The tensor is illustrated as a stacked set of matrices, where each slice corresponds to a 2D matrix, with 4 rows and 4 columns, indexed by the third index k. Each element in the tensor is denoted by the letter C plus three indices, i, j, and k. Here i represents the row index, j represents the column index, and k represents the slice number. While the k maximum value is fixed to 3, as it is an order-3 tensor, the maximum values of i and j are not constrained. To show this the second to last cells (in a row or column) contain three dots, and the last cell has m as a value for i, if it is the last row, and has as a value for j n, if it is the last column. M, and n, indicate the tensor’s dimensions in each direction. The second figure shows three tensors represented as circles. They have respectively one, two and three indices connected via a line to the circle, as they need one, two, and three indices to be accessed.}
    % \label{fig:enter-label}
\end{figure}

For example, a vector $A$ or a matrix $B$:
\begin{equation}
    A = \begin{pmatrix}
    A_1  \\
    A_2 \\
    \vdots \\
    A_m \\
    \end{pmatrix}
    B = \begin{pmatrix}
    B_{11} && B_{12} && \cdots && B_{1n} \\
     B_{11} && B_{22} && \cdots && B_{2n} \\
    \vdots &&   && \ddots && \vdots \\
    B_{m1} && B_{m2} && \cdots && B_{mn} \\
    \end{pmatrix} \\
    \label{eq:tensor-intro}
\end{equation}
are considered an order-1 tensor and an order-2 tensor, respectively while an order-3 tensor $C$ is represented as shown in Figure~\ref{fig:tensor-3d}.

A tensor network is a high-dimensional tensor that may be viewed as a graph where each node represents a tensor and the edges represent the connections between them.

A tensor of order-$k$ is an object with $k$ indices and can be represented with $k$ legs~\cite{tensor-review1,tensor-review2}, as shown in Figure~\ref{fig:tensor-legs}.

\begin{figure}[htbp]
    \centering
    \includegraphics[width=0.7\textwidth]{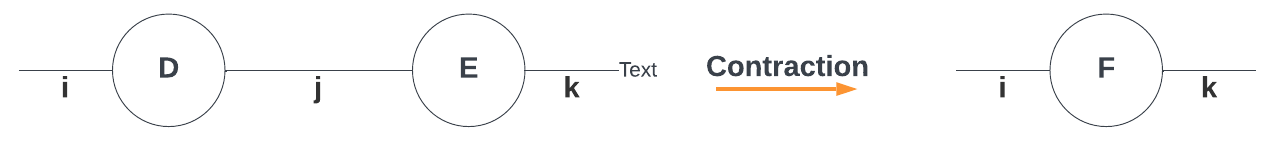}
    \caption{The tensors D and E are contracted into a single tensor F, with indices i and k.}
    \label{fig:tensor-contraction}
    \Description[The image shows a graphical representation of tensor contraction between two tensors, D and E, to produce a resulting tensor F.]{On the left side, there are two circles, labeled D and E, representing tensors. The tensor D is connected to index i on the left side and index j on the right side. The tensor E is connected to index j on the left and index k on the right. The shared index j between the tensors indicates that these two tensors are being contracted over this index. An orange arrow between the two sides, labeled "Contraction," indicates the operation of tensor contraction. The contraction over the shared index j results in a new tensor.
    On the right side, the contraction yields a single circle labeled F, representing the resulting tensor. This new tensor is connected only by the remaining indices i and k, since the index j has been summed over (contracted).}
\end{figure}

In order to work with tensors, the main operation is the tensor contraction, shown in Figure~\ref{fig:tensor-contraction}. Contracting two tensors means performing a summation over an index, which contracts the internal indices between the two tensors, mathematically represented as:
\begin{equation}
    F_{ik}=\sum_{j}{D_{ij}E_{jk}}
    \label{eq:tensor-contraction}
\end{equation}
The usual approach to handle the tensor networks is to contract the network from order $k$ to a single tensor. When the network contraction is divided in a sequence of binary contractions, the total computational cost is influenced by the choice of the contraction sequence. Working with pairwise contractions typically enables optimal computational performance, as they can be implemented as matrix-matrix multiplication~\cite{tensor-review1}.

% \section{Quantum Computers}
\section{Quantum computers}
\label{sec:qc}

Since the beginning of quantum research various algorithms\cite{quantum-algorithms-review} have been explored to verify if a quantum computer could solve some problems faster than classical machines.  
Examples of such quantum algorithms are Shor's algorithm~\cite{shor1999polynomial}, Grover's search algorithm~\cite{grover1996fast}, and quantum simulation~\cite{Nielsen_Chuang_2023, quantum-simulation}. 
Shor's algorithm, based on the quantum Fourier transform~\cite{Nielsen_Chuang_2023}, can be used to find the prime factors of an integer. Grover's search algorithm allows to speed up the search for an element which satisfies a certain known property. It allows in the case of a search space of size $N$ to find an element with no prior knowledge about the structure of the information in  $O(\sqrt{N})$ operations instead of the $O(N)$ operations required classically~\cite{Nielsen_Chuang_2023}. 
Quantum simulations are simulations of naturally occurring quantum mechanical systems, such as molecules,  using quantum computers~\cite{Nielsen_Chuang_2023}. \par
\begin{figure}[htbp]
\centerline{\includegraphics[width=0.5\textwidth]{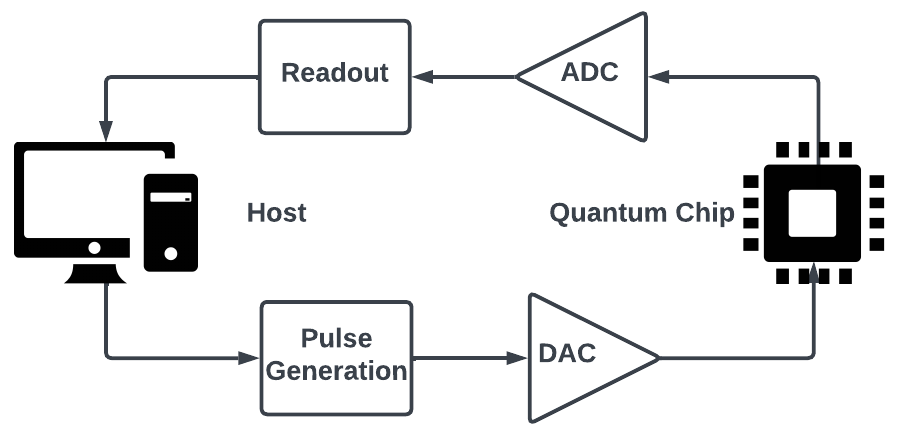}}
\caption{Basic scheme for a superconducting quantum computer and its control electronics. The control pulses generated are converted from digital to analog through the Digital-to-Analog Converter (DAC). The analog output of the quantum chip goes through an Analog-to-Digital Converter (ADC) and is then read.}
\Description[Quantum computer control electronics(host, pulse generation, readout), interface blocks (DAC, ADC) and quantum chip]{The image shows how the different parts of the system are connected. The different parts are placed in an elliptic configuration, starting from the host on the left and proceeding counter-clockwise back to the host. Starting from the host, represented as a computer symbol. An arrow goes from the host to the next component, the pulse generation block, which is represented as a rectangular box, situated on the bottom right of the host. The pulse generation block is then connected via an arrow to the base of the triangle used to represent the digital-to-analog converter. The point of the triangle, on the opposite side of the base, is on the right side, and it is connected with an arrow to the quantum chip, which is on the rightmost side of the figure. The quantum chip, represented as a chip symbol, is then connected to the analog-to-digital converter,  which is represented as the same triangle as the DAC,  but pointing in the opposite direction, and it is placed on the top left of the quantum chip. The ADC is then connected to the readout, represented as a rectangular box, and it is then connected back to the host. }
\label{fig:qc-stack}
\end{figure}

Quantum computers utilize different quantum chip technologies, including ion trap~\cite{ion-trap}, neutral atom~\cite{neutral-atom,Bluvstein2024}, semiconductor~\cite{semiconductor-arch}, and photonic~\cite{photonic,quantum-supremacy-photons1,Madsen2022}. This work focuses on the currently most common technology used to build quantum computers: the superconducting architecture~\cite{superconducting, superconducting-arch1,superconducting-arch2}. Figure~\ref{fig:qc-stack} shows the generic structure of a superconducting quantum computer.

\subsection{Quantum chip}
\label{subsec:quantum-chip}
The quantum chip, which includes the physical qubits, is the main component of the quantum computer.
 
Depending on the technology, they are controlled and connected in different ways. Connecting multiple qubits together allows for more complex circuits, comprising single- and multi-qubit gates.

As already mentioned in the introduction, one of the main limitations of current physical qubits is their short coherence time. This is the timescale of the exponential decay of the qubit superposition state~\cite{coherence-time-gate-fidelity}. The correctness of quantum computers is usually measured in terms of the fidelity of the operations or the output state.

Coherence times vary depending on the technology and the ``quality'' of the qubit. A different metric that allows to compare different technologies and different approaches is the amount of gates that can be executed during the coherence time of a single qubit. In the case of the current superconducting qubit implementations, they have a coherence time ranging from hundreds of microseconds\cite{coherence-time1, coherence-time2, coherence-time3}, up to close to a millisecond~\cite{coherence-time-superconducting}. Consequently, the control system, responsible for providing pulses to activate the qubits and read out their state, must be fast enough to be able to control thousands of gates during this time. 

Although some technologies, such as photonics, allow the quantum chips to work at standard room temperature, most of the other approaches require the chips to be cooled down to a temperature close to absolute zero to avoid thermal noise. Therefore, the chips are usually are placed inside cryogenic chambers. This introduces additional challenges related to controlling and communicating with the chips from other components of the quantum computer. If placed inside, the components need to operate at the low temperatures of the chamber and are forced to dissipate minimal power, while if placed outside the chamber, they face a bottleneck in scaling due to the physical limit of input and output cables available. In additional to the physical number of cables, each cable also carries heat into the chamber, thus increasing the thermal noise~\cite{cables-issue}.

\subsection{Pulse generation}
The qubit driving mechanism may vary depending on the type of quantum computers we briefly discussed in the previous section. In the case of the superconducting architecture, qubits are controlled by radio-frequency signals. Each qubit is characterised by a slightly different resonance frequency, which allows to drive them individually~\cite{intel-chip, chalmers-chip, superconducting-review}. The specific resonance frequency can vary slightly in time due to interaction with impurities in the qubit environment, affecting the fidelity of operations on each qubit. Therefore, periodic recalibration of the system is necessary~\cite{calibrate1, calibrate2, calibrate3}. This recalibration process allows to update the correct resonance frequencies used to control the qubit by the pulse generation hardware. Scaling up the number of qubits requires additional inputs to the system, which, as previously mentioned, becomes a major challenge for technologies that require a cryogenic chamber for cooling.

\subsection{Readout}
After the execution of the quantum circuit, the value of all or some output qubit must be measured. The main challenge is that discrimination between zeros and ones is non-trivial, leading to the development of various readout techniques~\cite{readout1, readout2, superconducting-review, readout3,output-ampl1,output-ampl2,readout-ecc,chalmers-readout}. The approaches range from improved amplification of the output at the analog level~\cite{readout3, output-ampl1, output-ampl2} to the introduction of error correction for fault-tolerant quantum computing using higher-abstraction level techniques such as machine learning~\cite{readout-ecc}. The main issue, as in the case of the pulse generation, is related to managing the connection between the readout system and the quantum chip through the cryogenic chamber, which is one of the major bottlenecks in scaling up the system. 

\subsection{Compiler}
Mapping an algorithm to a quantum circuit is a non-trivial task. To do this, we use specific tools such as a compiler, which must  take multiple decisions. Ideally, the compiler would assign each qubit in the algorithm to a physical qubit in the quantum computer, without considering the different error probabilities of each qubit. In this case, all the multi-qubit gate operations would occur between neighbouring physical qubits, giving the compiler freedom to choose any quantum gate operation.

In a real quantum computer, this is not always possible. First, different qubits on the same chip could have different probabilities of error due to the influence of the external environment. When reading the output of a qubit, the output might not always be correct, due to factors such as drifts in calibration, temperature variations, or measurement errors. Therefore mapping the operations to qubits according to their probability of error might increase the fidelity. 
Additionally, in some architectures, such as superconducting circuits, qubits can only interact with neighbouring qubits, requiring the addition of SWAP operations when direct interaction is not possible~\cite{mapping7-zhang,Siraichi2018,Paler2021,Yan2024}. 

Not every gate might is available on every quantum computer, and the choice of gate can influence the output fidelity.
Multiple approaches~\cite{mapping1-liu, mapping2-wang, mapping3-chen, mapping4-das, mapping4-wu, mapping5-patel, mapping7-zhang,Li2019,Zou2024,Cowtan2019,Sivarajah2021} have been proposed to address this issue, aiming to optimize the mapping of quantum circuit to the available qubits and gates. 

Thus, it is very important that the compiler considers all the characteristics of the available system as a sub-optimal compilation phase may result in resource under-utilization, high error rate, and lower fidelity~\cite{mapping1-liu}.

% \section{Quantum Computer Simulators}
\section{Simulations of quantum computers}
\label{sec:qc-sim}

In this section, we focus on how we can simulate, and optimize the simulation of, the execution of quantum computers. Simulations are becoming an increasingly valuable tool for evaluating and developing various components of the quantum computer stack. As in classical computing, each part of the system requires a distinct approach and level of detail.

\begin{figure}[htbp]
    \centering
\includegraphics[width=0.5\textwidth]{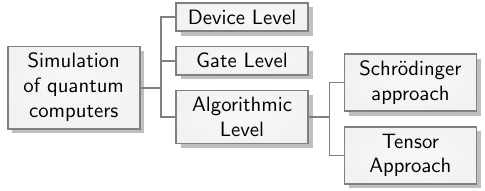}
    \caption{Taxonomy of simulation of quantum computers.}
    \Description[Diagram showing the taxonomy through a three representation, first branches into three different levels and then one branch is divided again in two approaches]{
    The diagram shows the different levels of quantum computer simulation. At the top level is "Simulation of quantum computers," which branches into three distinct levels: device level, gate level, and algorithmic level: this level is further subdivided into: Schrödinger Approach and Tensor Approach.
    }
    \label{fig:taxonomy}
\end{figure}

Simulations of quantum computers can be classified into the following three main categories, as also shown in Figure \ref{fig:taxonomy}.
\begin{itemize}
    \item {\bf Device level} --- this is the lowest level of the system and focuses on the materials and the implementation of the qubit. This is comparable to the low-level classical hardware simulation.
    \item {\bf Gate level} --- this is the middle level, where gates are mapped to qubits. This is comparable to the micro-architecture-level classical simulation.
    \item {\bf Algorithmic level} --- this is the highest level of the system, where algorithms are mapped to quantum circuits. This is comparable to the functional-level classical simulation. 
\end{itemize}

Although it is possible to simulate a complete quantum computer for a small number of qubits, this does not scale well for increasing number of qubits. This issue arises at every level and it is better described in each subsection. In general, the simulation execution is both memory- and compute-bound. With scalability in mind, simply trading one for the other is insufficient to achieve the desired performance, as beyond a certain number of qubits, one of the two limitations will become impossible to overcome. \par
We should notify the reader that in this work, we will not focus on the acceleration of device-level or gate-level simulations, as there are already plenty of solutions ~\cite{fem1, fem2, fem3, fem4}. Instead, we will focus on the research field of algorithmic-level simulation.

\subsection{Device level}
\label{subsec:device-level}
Device-level simulations help address the mechanical, thermal, and electromagnetic aspects of the problem. In the case of superconducting circuits, some mainstream commercial tools are reported here:
\begin{itemize}
    \item Ansys Electronic Desktop\cite{ansys} solves 3D Maxwell equations, and eigen-equations with various boundary conditions, including the option for lumped-element boundary conditions. This is useful for nearly all problems in the electromagnetic domain at RF and microwave frequencies.
    \item COMSOL Mulyiphysics\cite{comsol} is a Finite Elements Method (FEM) solver of many partial differential equations of physics. It is useful for simulating thermal, mechanical, and electromagnetic problems. London’s equations can be solved to simulate the Meissner effect in packages, chips, resonators, as well as effects of kinetic inductance. Heating or cooling of circuits and the impact of mechanical strain on packages or Printed Circuit Boards(PCBs) can also be simulated. 

    \item SolidWorks\cite{SolidWorks_2005} can be used for designing of mechanical parts such as nuts and bolts, packages, fixtures, heat sinks, and connectors. Designs can be exported and re-imported in Ansys tools for electromagnetic simulations. It is also possible to export the FEM mesh for use in other FEM-based simulators. FreeCAD is an open-source (python-based) version of this tool.
    \item InductEX\cite{InductEX}  from SunMagnetics Inc. solves London’s equations using FEM and is primarily used for extracting inductances in the layout of superconducting circuits, such as Rapid Single Flux Quantum (RSFQ) logic gates. 
\end{itemize}

Different types of simulators will be used in quantum dot technology, such as Intel Quantum Dot Simulator~\cite{IntelQuantumSDK} and nanoHUB Quantum Dot Lab~\cite{nanohub1,nanohub2}. Although related to different technologies, the simulators can be classified at the same device level. They evaluate the selected material and geometry and calculate information such as 3D visualization of the confined wave functions, incident light angle and polarization, and isotropic optical proprieties~\cite{nanohub1,nanohub2}.

\subsection{Gate level}
\label{subsec:gate-level}

Gate-level simulation focuses on the interaction between a few qubits connected by couplers, which are used to implement quantum gates. Evaluating different configurations is necessary to study and define the actual gates that will be used in quantum circuits. 

\begin{figure}[htbp]
\centerline{\includegraphics[width=0.5\textwidth]{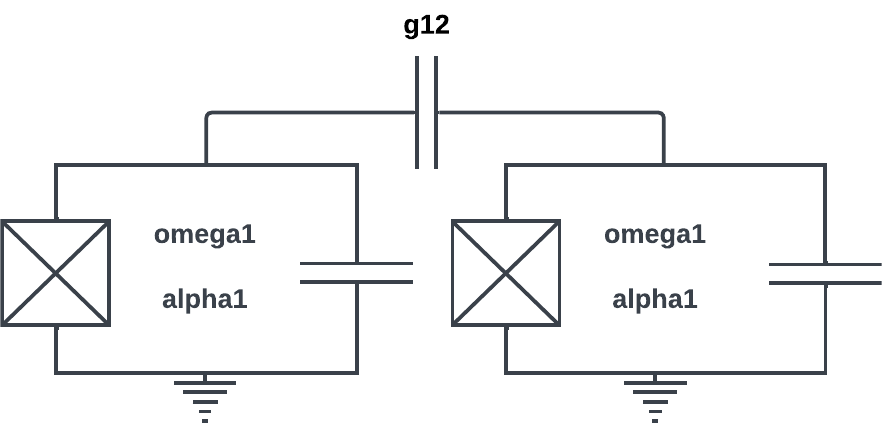}}
\caption{Simple circuit used for gate-level simulations, composed of two qubits and one coupler to connect them. Each qubit is modeled as a Josephson junction and a capacitor. In the circuit simulations a capacitive coupling is used to connect the two qubits~\cite{csqr, Fors2024}.}
\Description[The figure represents two coupled qubits, each depicted as a Josephson junction and a capacitance, and connected between them by a capacitor.]{Each qubit is shown as a block with the labels omega1 or omega2 and alpha1 or alpha2. Each block has a Jhosepson junction on the left, represented as a square box with lines crossing from corner to corner, and a capacitor on the right, represented with the standard symbol. On the bottom the box is shown as connected to ground and on the top is connected to the other box through a capacitive coupling, shown as a capacitor labeled g12. The qubits are placed one on the left and one on the right, with the capacitive coupling in the top middle of the figure.}
\label{fig:csqr}
\end{figure}

An example circuit used for gate-level simulation is shown in Figure 8. 
Simulations at this level are based on solving differential equations derived from the Schrödinger equation~\cite{qutip2}. 

There are different types of simulations at this level. %, mainly grouped into static and dynamic. 
Static gate-level simulation is useful for evaluating the energy levels of qubits and couplers, as well as assessing the interference between them. Dynamic gate-level simulation helps evaluate the effects of a pulse used to control the gate. This is crucial for determining the type and frequency of pulse needed on the quantum chip to ensure each qubit responds as expected.

Examples of simulations at this level are CSQR~\cite{csqr,Fors2024} and scQubits~\cite{scqubits1}\cite{scqubits2}. The gate-level simulation acceleration is based on the optimisation of the differential equation solution. 
Other simulators, such as QuTiP~\cite{qutip2}\cite{qutip1}, allow for run-time optimisation of the type of solver and hardware configuration, enabling faster simulations or reduced memory usage. Dynamiqs~\cite{dynamiqs} is another simulator that can run on CPU, GPU, and TPU, offering speedup through batching and parallelization on CPUs and GPUs. 
Another simulator at this level is QuantumOptics.jl~\cite{qojulia}, but no information on the acceleration is available.

\subsection{Algorithmic level}
\label{subsec:alg-level}
Algorithmic-level simulation is useful for testing the correct functioning of potential quantum algorithms.

\begin{figure}[htbp]
    \centerline{
    \includegraphics[width=0.4\textwidth]{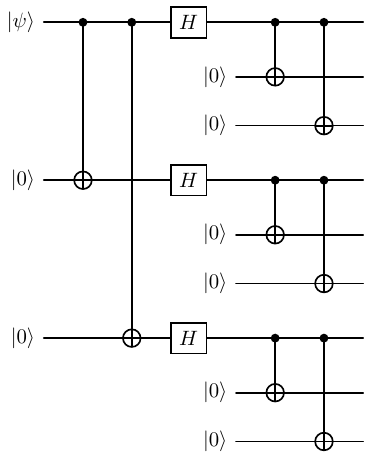}
    }
    \caption{9-qubit Shor's code circuit for error correction~\cite{shor-code}. Different gates are used in this circuit: multiple Hadamard gates which change the qubit state to the superposition state, and CNOT gates~\cite{Nielsen_Chuang_2023}.} 
    \Description[Multi-qubit quantum circuit, with nine qubits, three Hadamard gates and six CNOT gates]{
    From left to right, we have 9 qubits time evolutions represented as lines, that evolve in time from left to right. All qubits apart the first, psi, are initialized to the quantum state value zero On the first column we have the first qubit, psi, and two additional qubits, qubit four and qubit seven. In the first timestep, the starting qubit psi value is used to control through a CNOT gate the fourth qubit. On the second timestep, psi is used again to control through a CNOT gate the seventh qubit. On the third timestep an Hadamard gate is applied to psi, to the fourth and to the seventh qubit. In the next timesteps Psi, the fourth and the seventh qubit are used to control respectively the CNOT gate applied to the second, fifth and sixth qubit first, and in the last timestep they are used again to control the third, fifth and ninth qubit.
    }
    \label{fig:alg-level}
\end{figure}

One such algorithm is Shor's 9-qubit error-correction code~\cite{shor1999polynomial} shown in Figure~\ref{fig:alg-level}. Unlike device-level simulations, algorithmic-level simulations generally treat the quantum circuit as a sequence of matrix operations~\cite{Nielsen_Chuang_2023}.
Simulations may include noise, which models external factors such as temperature, electrical interference and additionally device changes in time, that may affect the state of the qubits. Noisy quantum states are represented as a density matrix, which requires more storage than the vector of a pure quantum state. Consequently, noiseless simulations need less memory and compute time, but they are farther from real-world conditions.
Nowadays, we are moving towards Noisy Intermediate-Scale Quantum computing (NISQ)~\cite{nisq,nisq-original}. Introducing noise in the simulations can provide a better correlation between simulation and real-world results. Below is a list of some of the most common algorithmic-level quantum circuit simulation tools available for use:

\begin{itemize}
    \item Intel Quantum SDK~\cite{IntelQuantumSDK}: Provides two different simulators with two different modes of simulation, Generic Qubits and Intel Hardware. The Generic Qubits simulation is a state-vector-based simulation, which uses the Intel Quantum Simulator as a backend, allowing for a qubit-agnostic execution. The Intel Hardware mode uses as its core the Quantum Dot Simulator, which simulates the Intel quantum hardware, currently under development. 
    \item QuEST~\cite{QuEST}: Simulator using state vectors and density matrices~\cite{Nielsen_Chuang_2023} for quantum circuits, which uses multithreading, GPU acceleration, and distribution to optimise the execution on multiple types of devices such as laptops, desktops, and networked supercomputers. 
    \item Qiskit~\cite{Qiskit}: Open-source SDK for working with quantum computers. Used as the core library for different types of simulations. It allows both for noiseless and exact noisy simulation with Qiskit Aer~\cite{Qiskit}.
    % An example of such circuit is the Shor algorithm \cite{shor1999polynomial} for the factorization.
    \item HyQuas~\cite{HyQuas}: This is a hybrid partitioner-based quantum circuit simulator, optimized for GPUs. It selects the optimal simulation method for different parts of a given quantum circuit. It implements additional solving methods and makes use of distributed simulation.
    \item qTask~\cite{qtask}: Quantum circuit simulator that focuses on optimizing the speed in case of modifications of a small part of the circuit, defined as incremental simulation. When this happens, the state amplitudes (probability amplitudes of the state) are incrementally updated, removing the need for simulating the whole circuit every time. 
    \item PennyLane~\cite{bergholm2022pennylane}: Open-source software framework for quantum machine learning, quantum chemistry, and quantum computing. Supports GPU acceleration by making use of NVIDIA cuQuantum SDK~\cite{bayraktar2023cuquantum}.
    \item QuTip-qip~\cite{qutip-qip}: QuTiP quantum information processing package, which offers two approaches to quantum circuit simulation. The algorithmic-level approach is based on the circuit evolution under quantum gates by matrix multiplication, while another approach uses open system solvers in QuTiP~\cite{qutip1, qutip2} to simulate noisy quantum devices.
\end{itemize}

While gate-level simulations focus on the full time evolution from the start to the end of the gate, algorithmic-level simulations consider only the changes that the gate's corresponding matrix applies to the state vector. Although these simulators provide less detail, they allow for faster simulations or alternatively an increased number of simulated qubits. However, they are also constrained by memory and time. The memory required to simulate a circuit is determined by the number of qubits, the number of gates, the data representation, and the number of times the circuit is executed, which, in the case of a real quantum circuit, corresponds to the number of measurements.

Most simulators are designed for algorithmic-level simulation, which is the furthest from actual physical implementation and allows for evaluating problems from a hardware-agnostic perspective. Quantum simulations at the algorithmic level can be categorized into three main types of approaches:
\begin{itemize}
    \item Schrödinger-style simulation~\cite{Nielsen_Chuang_2023}
    \item Feynman-style simulation~\cite{feynman-based-simulation}
    \item Tensor-based simulation~\cite{tensor-network-simulations}
\end{itemize}

While the Schrödinger and Feynman approaches have been the primary methods in the past, there is now greater focus on exploring the potential of the tensor-based approach.

\subsubsection{Schrödinger-style simulation}
Schrödinger-style simulation, also known as state-vector-based simulation, is the mainstream technique for general-case simulation of quantum algorithms, circuits, and physical devices. As described earlier, a quantum state is represented by a vector of complex-valued amplitudes, and the primary approach in this simulation is to store the current state vector and iteratively multiply it by a state transformation matrix~\cite{qtask}.

Schrödinger-style simulation resources scale linearly as a function of the circuit depth and exponentially as a function of the number of qubits. Additionally, it allows for relatively straightforward implementations that are commonly used for small and mid-size quantum circuits and device/technology simulations~\cite{fatima}.

\subsubsection{Feynman-style simulation}
Feynman-style path summation considers each gate connecting two or more qubits in a quantum circuit as a decision point from which the simulation branches. The final quantum state is obtained by summing up the contributions of the results of each branch, which are calculated independently. In comparison to Schrödinger-style simulations, traditional Feynman-style path summation~\cite{feynman} uses very small amounts of memory but doubles the runtime on every (branching) gate. This results in a much longer runtime and does not allow for optimal memory usage, as the number of paths grows exponentially as a function of the decision points. 
Unlike traditional Schrödinger-style simulation, Feynman's resulting algorithms are depth-limited, making them a good fit for near-term quantum computers that rely on noisy gates~\cite{Markov}.

\subsubsection{Schrödinger-Feynman hybrids}
A possible approach to leverage both Schrödinger and Feynman approach is Schrödinger-Feynman hybrids~\cite{feynman}\cite{feyn-scho2}, proposed in the work of Markov et al.~\cite{Markov}. In the context of nearest-neighbour quantum architectures~\cite{near-neighb-qarch1, Arute2019, Kim2023}, where the qubit array is partitioned into sub-arrays, the Schrödinger approach is applied to each sub-array. This reduces the memory requirements for $k$ qubits from $2^k$ to $2^{\frac{k}{2}+1}$, although this introduces a dependency on the number of gates acting across the partition. 

The gates are decomposed into a sum of separate terms to allow for independent simulation, but this also results in an increase in execution time. For example, in the case of a CZ gate, the gate can be decomposed into two terms, which results in doubling the run time. However, this increase in run time is still lower compared to the Feynman-style path summation described earlier.

\subsubsection{Tensor-based simulation}
The Schrödinger approach typically stores the full state of the qubits, allowing for the simulation of arbitrary circuits. However, the downside is the exponential increase in memory requirements. For an $n$-qubit system, $O(2^{n})$ space is needed.

When using the tensor-based approach, the quantum circuit is described as a tensor network, with each $n$-qubit gate represented as a rank-2$n$ tensor. This transforms the simulation into a problem of contracting the corresponding tensor network. Tensor network contraction is performed by convolving the tensors until only one vertex remains. For circuits with a large number of qubits and shallow depth (as complexity often grows exponentially with circuit depth), this method is highly efficient. Thus, using tensor networks allows for the simulation of only one or a small batch of state amplitudes at the end of the circuit. The complexity of the tensor network is constrained by the largest tensor involved in the contraction process.

% \section{Quantum Computer Simulators Acceleration}
% \label{quantum-sim-acc}
% \input{qc-sim-acc}

% \section{Acceleration Hardware Platforms}
\begin{figure*}
    \centering
    \includegraphics[width=\textwidth]{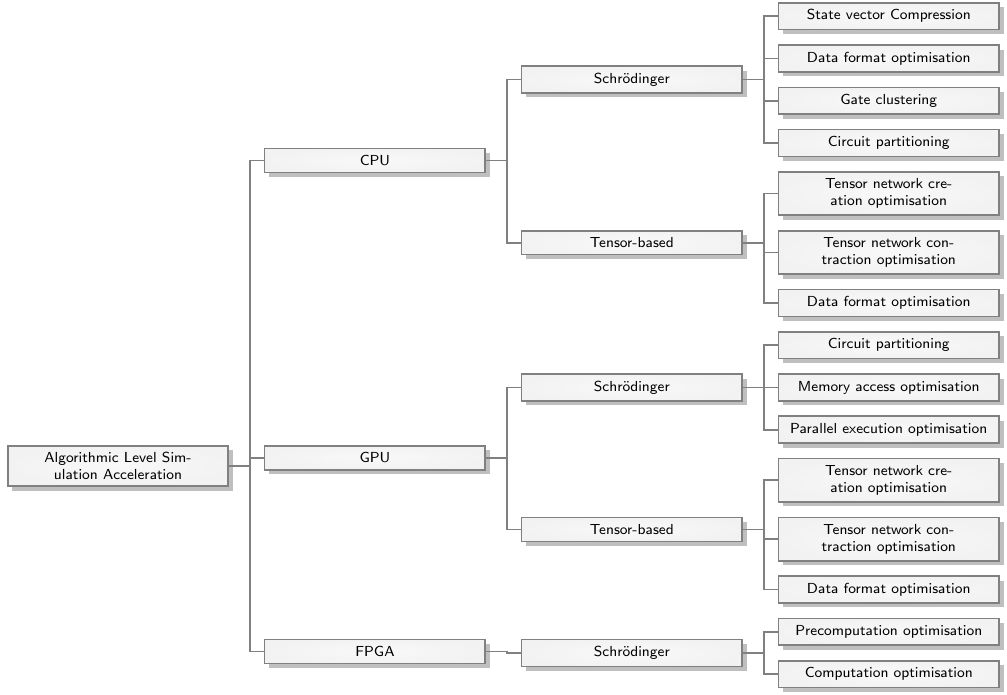}
    \caption{Summary of acceleration methods.}
    \label{fig:acc-meth-summary}
    \Description[Treee diagram with three branching levels, the first node branches into the hardware platforms, then into the general level approaches, then into the specific optimization categories]{
    The diagram shows a hierarchical structure of various hardware architectures and methods used for algorithmic-level simulation acceleration in quantum computing. It breaks down into three main hardware platforms: CPU, GPU and FPGA. Under CPU, two primary simulation methods are highlighted: Schrödinger and tensor-based. Schrödinger includes state vector compression, data format optimization, gate clustering, and circuit partitioning. The tensor-based include tensor network creation optimization, tensor network contraction optimization, data format optimization, and circuit partitioning. Under GPU, similar simulation methods are categorized under Schrödinger and tensor-based. Schrödinger techniques include data format optimization, Circuit partitioning, Memory access optimization, and parallel execution optimization. Tensor-based techniques include Tensor network creation optimization, Tensor network contraction optimization, and Data format optimization. Under FPGA, only Schrödinger-based methods are considered, including precomputation optimization and computation optimization.
    
    }
\end{figure*}

\section{Simulation Hardware Platforms}
\label{sec:sim-plat}

Based on the size of the problem, various hardware platforms with different computational power can be used to run the simulations. Nevertheless, all of them will eventually be limited by either the required memory or the simulation time. The exponential growth of memory and simulation time is the reason for the relevance of the simulation optimization efforts, which will be discussed later in Sections \ref{sec:cpu-acc} to \ref{sec:fpga-acc}.
A diagram presenting the techniques discussed in this work is depicted in Figure~\ref{fig:acc-meth-summary}.

\subsection{Small-scale quantum simulation}
Currently, an algorithmic-level simulation of 30 qubits with Intel quantum SDK~\cite{IntelQuantumSDK} requires up to 34GB of free RAM. Increasing this by just two additional qubits would require up to 135GB. This means that with today's technology, even a common personal computer can be used for a moderately small number of qubits. 

Personal computers are the most commonly available platform for running simulations, primarily using the CPU and, in some cases, the GPU. Personal computers can also be combined with an external FPGA, which allows off-loading some of the work. As described in Section IV-C, in state-vector simulations, the core operation is matrix multiplication, which, if properly handled due to the use of complex numbers, can be processed inside the FPGA and the results then transferred back to the CPU.

\subsection{Medium-scale quantum simulation}
Scaling is quite an issue in quantum computer simulations, as mentioned earlier. This increase in computational resources can be supported by better simulation hardware, such as workstations. These are typically equipped with state-of-the-art CPUs and one or more GPUs, offering more memory compared to a standard personal computer. This enables the simulation of a larger number of qubits, although still limited due to the exponential scaling of memory requirements. With current technology, simulating up to 32 qubits is possible using commercially available workstations~\cite{IntelQuantumSDK}. 
Another factor to consider is the simulation time. Upgrading to workstations, especially those equipped with multiple GPUs, generally enables faster execution if the simulator is optimized for parallel execution. Additional speedup can be achieved if the workstation has a processor with a matrix unit~\cite{matrix-unit}, which allows for quicker execution of core operations.

\subsection{Simulation beyond the medium-scale }
If the goal is to run simulations with the highest possible number of qubits or at the fastest speed possible, a high-performance computer is required. An example of this is the execution of 61-qubit quantum circuits~\cite{60qubit} on Argonne Theta~\cite{argonne_theta} using qHIPSTER~\cite{qhipster}, an earlier version of the Intel quantum simulator~\cite{IntelQuantumSDK}. The simulation, although optimized, required 0.8 Pb of memory, a level of resources not available on standard laptops or workstations. The advantage of high-performance computers is that they are equipped with multiple nodes, each composed of multiple CPU cores or GPUs, allowing for significantly better performance compared to a single laptop or workstation.

% % \section{Acceleration Hardware Platforms}
% \input{hw-independent}

% \section{CPU based hardware acceleration}

\section{Acceleration using CPU}
\label{sec:cpu-acc}

\begin{table*}[htbp]
\caption{Summary of most important  acceleration works with CPU.}
%* denotes results based on simulator developed by the authors}
\begin{center}
\resizebox{\textwidth}{!}{%
\begin{tabular}{ccccc}
\toprule
Work & Simulation type & Baseline & Improvement & Benchmark Platform \\
\midrule
\cite{Lykov} & Tensor-based & 120-qubit QAOA~\cite{baseline-lykov-2022} & 210-qubit QAOA in \qty{64}{s} & Theta Supercomputer \\
\cite{Yong}   & Tensor-based & 81-qubit RQC, 40 depth~\cite{baseline-yong-2021} & 100-qubit RQC, 42 depth in \qty{304}{s} & Sunway Supercomputer \\
\cite{Markov} & Schrödinger & 45 qubits, 26 depth, \qty{0.5}{PB} mem~\cite{haner2017} & 45 qubits, 27 depth, \qty{17.4}{GB} mem in \qty{1.4}{\hour} & Google Cloud \\
\cite{fatima} & Schrödinger & 36 qubits in \qty{194.12}{s}~\cite{qsim} & 36 qubits in \qty{94.48}{s} & Linux Server \\
\cite{li2020sunway} & Schrödinger & 49 qubits, 27 depth in $\geq$\qty{24}{\hour}~\cite{baseline-li-2020} & 49 qubits, 27 depth in \qty{1.49}{\hour} & Sunway \\
\cite{60qubit} & Schrödinger & 47 qubits, \qty{2.8}{PB} mem~\cite{Nielsen_Chuang_2023} & 61-qubit Grover's search, \qty{0.8}{PB} mem & Argonne Theta \\
\cite{deraedt2018} & Schrödinger & 45 qubits, JUQCS-E~\cite{deraedt2018} & 48 qubits in \qty{300}{\minute} & K Computer (RIKEN) \\
\bottomrule
\end{tabular}%
}

\label{tab:summary-cpu}
\end{center}
\end{table*}

CPU simulation acceleration focuses mainly on scalability, and in particular, most of the work reported here tries to optimize the simulation on big clusters of CPUs, with high-performance computers such as Summit~\cite{summit}, Sierra~\cite{sierra} (Lawrence Livermore National Laboratory), or Theta~\cite{argonne_theta} (Argonne National Laboratory). 
The optimizations presented are classified by their approach: Schrödinger or Tensor-based. 
Table~\ref{tab:summary-cpu} summarizes all the results.

\subsection{Schrödinger-style simulation acceleration}
As described previously in Section~\ref{subsec:alg-level}, accelerating  Schrödinger-style simulations consists of speeding up the core matrix multiplication. Different approaches have been taken to achieve improvements. 
\subsubsection{Baseline simulation algorithm}
The most mathematically simple way to execute full state-vector simulations for an $n$-qubits circuit with quantum gates, reported in the work of Fatima and Markov~\cite{fatima}, is to:
\begin{enumerate} 
    \item Order the gates left to right 
    \item Execute each layer of gates in parallel. All qubits must go through either an actual gate or an identity matrix; therefore pad each gate with an identity matrix of an appropriate dimension via Kronecker products to obtain a $2^n \times 2^n$ matrix
    \item Multiply all matrices in order
\end{enumerate}
This allows the entire circuit to be represented as a matrix, and multiplying it with the input state vector produces the output. Although it is a simple algorithm, it is not the most efficient method in terms of memory utilization and parallelizing computations. Another straightforward approach is to apply each gate directly to the input state vector. Generally, in most works, the simulations are performed in two main steps. The first step is to partition the circuit into multiple sub-circuits, which can be executed in parallel, eventually grouping or reordering gates within the same partition. The second step involves adopting various strategies to optimize the execution. Additional techniques, such as compression and optimizing the data format, can further enhance performance.

\subsubsection{State-vector compression}
One of the main issues of the Schrödinger simulation, mentioned already before, is memory usage. Storing the full state vectors during the simulation requires an exponentially increasing memory space. A possible solution, used already in other applications that handle vast volumes of data, is the use of data compression techniques. 

There are mainly two types of compressions: lossless compression~\cite{gzip, zstd, blosc} and error-bounded lossy compression~\cite{sz, isabela, fpzip, zfp, vapor, sasaki}. 
Introducing a module in the simulation workflow that compresses the state vectors before storing them in memory and decompresses them when needed helps address the memory usage problem.

Lossless compression is an approach used in works such as Zhao et al.~\cite{Zhao}. Their work focuses on reducing the data transfer between CPU and GPU, and one of their contribution is the lossless data compression of non-zero amplitudes, achieved by observing how the state vector after each operation has similar amplitude values. Lossy compression of state vectors is the approach used in works such as Wu et al.~\cite{60qubit}. That approach allows to simulate Grover's search algorithm, as well as other quantum algorithms such as Quantum Approximate Optimisation Algorithm (QAOA)~\cite{qaoa} and random circuit proposed by Google~\cite{random-quantum-circuit}, with up to 61 qubits. This is achieved by compressing the state vectors and reducing the memory size required from \qty{32}{EB} to \qty{768}{TB}. For the data compression, the authors implement an error-bounded lossy compressor, a technique that limits the compression ratio based on the corresponding fidelity loss. It is possible to select the optimal compression strategy during different parts of the simulation, using Zstd~\cite{zstd} alongside an error-bounded lossy compression method developed by the authors~\cite{60qubit}. 

Another approach, presented in~\cite{deraedt2018}, is to use an adaptive encoding scheme based on the polar representation $z=re^{i\theta}$ of the complex number $z$ for the state-vector elements. One byte is used to encode the angle $-\pi \leq \theta \leq \pi$ and the remaining bytes encode the value r, after scaling based on the maximum and minimum values of the state vector. This allows for a reduction in the amount of memory required to store the state by a factor of 8, but it requires additional time to perform the encoding and decoding procedure, up to a factor of 3-4.

\subsubsection{Data format optimisation}
While the 64-bit double representation is the most common approach to the data representation~\cite{deraedt2018}, there have been efforts to reduce the number of bits required to obtain results with a good enough fidelity. This allows for a lower memory footprint and higher throughput. 

The work of Fatima and Markov~\cite{fatima} uses the 32-bit float type representation instead of the more common 64-bit double type, therefore reducing the total amount of necessary resources. The approximation error is controlled by globally keeping track of the changes in amplitude for changes bigger than $\frac{1}{2}$. 

This is done because most of the gates do not significantly change the amplitude value. Gates more commonly act on the phase, so the result is not at risk of underflow. An underflow occurs when the result of an arithmetic operation is relatively so small that it can not be stored in the input operand format without resulting in a rounding error that is larger than usual.

But some gates, such as the Hadamard gates, can change the amplitude by a factor of $\frac{1}{2}$ or $\frac{1}{\sqrt{2}}$, generating underflow issues with the 32-bit float representation. In this case, the change is tracked and included at the end of the quantum circuit, allowing the reduction of memory needed while still obtaining results with good fidelity. 

The same research group published a follow-up work~\cite{Markov} where their simulation acceleration techniques are optimized in terms of parallelization, with a focus on allowing the simulation to be run on generic hardware.

\subsubsection{Gate clustering and circuit partitioning}
Simulating gates one at a time, as described in the baseline algorithm, is slow because it requires separate memory traversals~\cite{fatima}. A common technique to avoid this issue is to simulate gate clusters in batches. A gate cluster is obtained by combining the matrices representing the gate matrices acting on one or multiple qubits. The common approach is to cluster adjacent gates acting on the same qubit. Google QSim~\cite{qsim} merges each one-qubit gate to some nearby two-qubit gate. Another work~\cite{haner2017} applies gate clustering to a larger amount of qubits, up to five, and then multiplies out the obtained gate matrices.

The work of Fatima and Markov~\cite{fatima} creates clusters of $q$ qubits out of the total amount of $n$ qubits that grows as $O(q^2)$. They apply gate clustering by reordering, a technique that loops over the circuit and clusters adjacent gates of the same type or reorders non-adjacent gates to form larger clusters, allowing for optimized algorithms for each type of cluster. The approach of qTask~\cite{qtask} is to partition a state vector into a set of blocks, with each partition spawning one or multiple tasks performing gate operations on designated memory regions. This, as the previous techniques, allows to enable inter-gate operation parallelism due to the breaking down of the gate dependencies.

Another strategy presented in ~\cite{li2020sunway} is the technique named implicit decomposition. The target circuit is efficiently partitioned into different parts with a focus on balancing the memory requirements for each one of them. This efficient partitioning allows to save memory space compared to storing the entire state vector, because it is not necessary to have all the amplitudes after the individual calculations.
%, but only a subset of them. 
They additionally propose a dynamic algorithm to select the optimal partition scheme.

\subsubsection{Memory access optimization} 
Given that the execution is dominated by matrix multiplication operations, an obvious optimization is to do them in a

cache-friendly way, resulting in improved performance~\cite{fang2022}. 
Memory locality can be exploited with gate clustering. If paired-up gates act on qubits as closely as possible, it reduces memory strides (distance between two successive elements of an array in memory) when simulating gate pairs acting on less significant bits~\cite{fatima}. 

Another technique, described in~\cite{fatima}, is cache blocking, where rather than applying pairs of qubit gates in separate passes, such pairs, acting on different qubits, are reordered and applied partially in different orders. Each state vector is divided into chunks, which fit in the L2 cache, and for each chunk, multiple non-overlapping pairs of one-qubit gates and an occasional unpaired gate are applied to it. This reduces the cache misses and improves the performance. The work of Fang et al.~\cite{fang2022} tackles the memory locality with their hierarchical simulation framework. The authors consider that scaling up means the working set size of the simulation set would exceed the cache size of a modern CPU. To address this, the input circuit is executed as a sequence of sub-circuits, each containing a portion of the original gates, allowing for better locality.

\subsubsection{Parallel execution}
Efficiently handling the parallel execution of different gates or tasks is crucial to capitalizing on the advantages introduced by the circuit modifications mentioned earlier. Most works utilize a multi-CPU approach with OpenMP and MPI-based CPU simulations. The work of Fatima and Markov~\cite{fatima} explains in detail how to exploit CPU architecture to simulate different clusters, exploring data-level parallelism.

\subsection{Tensor-based simulation acceleration}

Tensor network-based simulations are divided into two main steps. The first one is to represent the quantum circuit as a tensor network. The second step is to apply the contraction to the tensors, as described previously in Section~\ref{subsec:tensor-networks-bg}. We now present the different acceleration solutions for each step of the simulation. 

\subsubsection{Tensor network creation}
For the first step of the simulation,
different approaches can be taken depending on the type of algorithm and the structure of the quantum circuit. 
In the case of Random Quantum Circuits (RQC), a viable solution is using Projected Entangled Pair States (PEPS) representation of quantum states from many-body quantum physics~\cite{Yong}. This representation, according to \cite{Yong} works for $2N$ by $2N$ lattice types, but to work with different structures, the generation of the best contraction path (choice of which nodes to contract at different steps of the simulation) becomes a bigger issue. A possible approach is to use the CoTenGra software~\cite{CoTenGra} to look for the best path. It uses a loss function that combines the considerations for both the computational complexity and the compute density, which are determinant factors for the performance on a many-core processor \cite{Yong}. Another approach is to use ordering algorithms. There are several available ordering algorithms, classified as:
\begin{itemize}
    \item Greedy algorithm: usually used as a baseline, contracts the lowest-degree vertex in a graph.
    \item Randomized greedy algorithms: The computational cost of the contraction at each step is function of the maximum number of neighbours for a node. Minimizing this by choosing the optimal contraction order allows for a decreased computation cost. This approach can do so without prolonging the run time \cite{Lykov}.
    \item Heuristic solvers: Attempt to use some global information in the ordering problems. Example of this are QuickBB\cite{QuickBB} and Tamaki's heuristic solver\cite{tamaki2018positiveinstance}.
\end{itemize}

 \begin{table*}[htbp]
\caption{Summary of most important works acceleration performances with GPU.}
\begin{center}
\resizebox{\textwidth}{!}{%
\begin{tabular}{ccccc}
\toprule
Work & Simulation type & Baseline & Improvement & Benchmark Platform \\
\midrule
\cite{bayraktar2023cuquantum} & Multiple  & 53-qubit RQC, depth 10~\cite{pytorch,cupy,numpy} & \makecell{Average contraction speedup: \\ $4.05\times$ to~\cite{pytorch}, \\ $4\times$ to \cite{cupy}, \\ $547.35\times$ to~\cite{numpy}} & \makecell{NVIDIA H100 GPU, \\ NVIDIA A100 GPU, \\ AMD EPYC 7742 CPU} \\
\cite{willsch2022}  & Tensor based & 42 qubits in \qty{1965.52}{s}~\cite{deraedt2018} & 42 qubits in \qty{195,71}{\s} & NVIDIA A100 GPU\\
\cite{lykov2021}  & Tensor based & 30 qubits, depth 4 in $246 \, s$~\cite{numpy} & 30 qubits, depth 4 in \qty{1,4}{\s} & GPU \\
\cite{fang2022}  & Schrödinger & 30-37 qubits circuits in range \qtyrange{10}{100}{\s}~\cite{iqs} & \qty{15.8}{\percent} average runtime reduction & NVIDIA V100 GPU \\
\cite{doi2020}  & Schrödinger & 35-qubit QFT & 35-qubit QFT, \qty{80}{\percent} speedup~\cite{baseline-doi-2020} & $6\times$ NVIDIA V100 GPU\\
\cite{Li2020sc20}  & Schrödinger & 26 qubits in less than \qty{10}{\ms}~\cite{baseline-li-2020-sc20, baseline-li-2020-sc20-2} & 26 qubits, $\geq10\times$ speedup & GPU Cluster \\
\cite{doi2023efficient}  & Schrödinger & 22-qubit QFT~\cite{Qiskit} & Noisy QFT, 22 qubits, up to $10 \times$ speedup & GPU \\
\bottomrule
\end{tabular}%
}

\label{tab:summary-gpu}
\end{center}
\end{table*}
 
\subsubsection{Tensor-network contraction}
Depending on the optimisations done during the tensor creation step, the contraction can be more or less efficient. During the contraction phase, an optimal slicing method is required to divide the tensor network into different clusters. This allows an efficient parallelization of the computations, to efficiently process all the sub-tasks of the circuit across all the available nodes. This is necessary to balance the compute and storage costs. In the graph representation, the contraction of the full expression is done by consecutive elimination of graph vertices. The consequence is that the vertex is removed from the graph and the neighbours are connected~\cite{Lykov}.  The slicing of a tensor over an index means evaluating many variables while keeping one of them constant. Finding the optimal index to slice is the focus of this process. An approach to this is to select the best index after each step of the slicing, a method known as step-dependent slicing~\cite{Lykov}. \par
Another approach is to divide the contraction into two steps, the index permutation of the tensors as a preparatory step and the second step is the following matrix multiplication to achieve the contraction results~\cite{Yong}. Permutation of indices is generally required to convert the tensor contractions into efficient matrix multiplications. In the case of tensor contraction on many-core processors with a high compute density of high-rank tensors, it is important to reduce the permutation cost to reduce the movement of data items with strides in between, which is unfriendly for current memory systems~\cite{Yong}\cite{villalonga2020}. A proposed approach is to use fused permutation and multiplication, which means using different compute processing elements in a collaborative way~\cite{Yong}, where each CPU reads its corresponding data block in a regular, constant stride, pattern, thus achieving a high utilization of the memory bandwidth. 
Parallelization of the tensor computation is done after the slicing. A possible approach is to use a two-level parallelization architecture. On a multi-node level, the partially contracted full expression is sliced over $n$ indices and distributed to $2^{n}$ message-passing interface ranks. Node-level parallelism over CPU cores is done using system threads. For every tensor multiplication and summation, the input and output tensors are sliced over $t$ indices. The contraction is then performed by $2^t$ threads writing results to a shared result tensor~\cite{Lykov}.

%Could be more about the actual permutation 

\subsubsection{Data format optimisation}
To achieve a high accuracy in the computation of the simulated output state compared to the actual output state it is necessary to represent the data with a correct amount of bits. The default approach is to use 64-bit double (floating-point) representation~\cite{deraedt2018}. In the case of RQCs, the work of Yong et al.~\cite{Yong} proposes an adaptive precision scaling, which adjusts the data precision to single or half-precision dynamically depending on the degree of the sensitivity of different parts of the computation.

\subsection{Other acceleration approaches}
Alternatively to what was described previously, some works exploit other approaches for optimizations, such as Quantum Processing Units (QPUs)~\cite{willsch2022}\cite{Tang} and different memory technologies. For example, CutQC~\cite{Tang} employ classical CPUs interacting with a small quantum computer. This hybrid computing approach enables the evaluation of larger quantum circuits that cannot be executed on a classical computer alone. It offers better fidelity and allows the simulation of larger circuits than either approach could achieve independently. SnuQS~\cite{SnuQS} focuses on the full utilization of the storage device connected to the system. That work aims to achieve maximum I/O bandwidth by employing memory management and optimization techniques, resulting in better performance compared to DDR4 DRAM main-memory-only systems. Other approaches that make use of multiple hardware platforms can be found in the literature~\cite{villalonga2020, zhang2023hqsim,efthymiou2021, Efthymiou2022quant-sim-just-in-time, dou2022qpanda, doi2019quantcompsim-on-hpc, deavila2023stateoftheartquantcompsim, li2021sv-sim}

% \section{GPU based hardware acceleration}

\section{Acceleration using GPU}
\label{sec:gpu-acc}

 Accelerating quantum computer simulations using GPUs has been an effective approach for many years. Early works, such as the one by Gutierrez et al.~\cite{gutierrez2010}, attempted this over 10 years ago. Similar to CPU simulations, GPU simulations can also be classified according to the classification defined in Section~\ref{sec:qc-sim}. The results from the most relevant works are presented in Table~\ref{tab:summary-gpu}.

\subsection{Schrödinger-style simulation acceleration}
As seen in Section~\ref{sec:cpu-acc}, the core operations of quantum circuit simulation are small matrix multiplications. If the data dependencies between these operations are limited or nonexistent, and they can be parallelized, GPUs can perform them with much higher throughput compared to CPUs. It is still important, even for GPU execution, to focus on data locality to minimize memory access misses~\cite{HyQuas}. In GPU simulation acceleration, each work focuses on one or both of two steps: first, partitioning the circuit into sub-circuits or tasks; second, accelerating the computation.

\subsubsection{Circuit Partitioning}
As seen for the CPU in Section~\ref{sec:cpu-acc}, also in the GPU approach the first step followed is to partition the circuit in order to enable a more efficient parallel execution. HyQuas~\cite{HyQuas} proposes a circuit-aware partition strategy and a high-accuracy performance model that guides the partitioning. This allows to obtain a near-optimal partition of a given quantum circuit into different groups and select an optimal method to compute each one. CuStateVec, a library of the cuQuantum SDK~\cite{bayraktar2023cuquantum}, which focuses on state-vector simulation acceleration for GPU, uses gate fusion~\cite{qhipster}. Numerous small gate matrices are fused into a single multi-qubit gate matrix, which can then be computed in one shot instead of performing multiple computations. This allows for improved performance in cases where both the high compute performance and high memory bandwidth of the GPU are used. %Diagonal fuse matrices, in practice it's optimized for diagonal ones 

\subsubsection{Memory access optimisation}
Different approaches can be used to address the data locality issue. ShareMem~\cite{sharemem} method considers a circuit partitioned in several gate groups, which are applied only to a subset of $k$ qubits out of the $n$ total qubits. The total $2^n$ state vector values can be split into fragments of size $2^k$, which can be stored in GPU shared memory, and each fragment can be mapped to one GPU thread block. The thread block can load the fragment from the global memory to the shared memory, apply the gates on it, and store back the fragment in the global memory. This method performs better than the following BatchMV approach in the case of a sparse part of the circuit~\cite{HyQuas}.

The BatchMV~\cite{batchmv} method is based on the idea that if all the target and control qubits of a gate group come from the same $k$ qubits, it is possible to merge these gates into a $k$-qubit gate. This allows to divide the simulation into $2^{n-k}$ matrix-vector multiplication tasks, each one with a $2^k\times 2^k$ gate matrix to a state vector of $2^k$ values. The index of the values only differs on these $k$ positions, improving the data locality. This method performs better than the previous in the dense part of the circuit. HyQuas~\cite{HyQuas} automatically selects the best approach depending on the part of the circuit between OShareMem, an improved version of ShareMem~\cite{sharemem} and TransMM, which performs better than BatchMV in their framework. TransMM transposes the quantum state, allowing to treat the gate-applying operation into a standard GEneral Matrix Multiplication (GEMM) operation, which can be accelerated by highly optimized libraries such as cuBLAS and Tensor Cores.

\subsubsection{Parallel execution}
As seen in Section~\ref{sec:cpu-acc}, the main way to approach distributed computing is with OpenMP or MPI. 
It is important to maximize GPU usage while minimizing data exchange between the CPU and GPU, as frequent amplitude exchanges between them introduce significant data movement and synchronization overhead~\cite{Zhao}. 
Qubit reordering~\cite{deraedt2007} is also a technique used by cuStateVec~\cite{bayraktar2023cuquantum} to address the challenge of applying a gate onto a global indexed qubit. The approach involves moving the qubit from a global index to a local index, allowing a single GPU to compute the state vector. On a more broad approach, there are even works proposing a novel multi-GPU programming methodology, such as the work from Li et al.~\cite{Li2020sc20}, which constructs a virtual BSP machine on top of modern multi-GPU platforms.

\subsection{Tensor-based simulation acceleration}
As reported in the work of Vincent et al.~\cite{Vincent2022jetfastquantum} a major issue in quantum computer simulations is the under-utilization of GPUs and CPUs in supercomputers. They further predict an increase in inefficiencies due to the increase of parallelism and heterogeneity in exascale computers, with billions of threads running concurrently~\cite{inefficiency-cpu-gpu,inefficiency-cpu-gpu2}. Future supercomputers will see an increase in on-node concurrency rather than the number of nodes, with large multi-core CPUs and multiple GPUs per node~\cite{exascale}.

\subsubsection{Tensor network creation}
For this step of the simulation on GPUs, some works~\cite{Vincent2022jetfastquantum} also use CoTenGra~\cite{CoTenGra}, as already mentioned in Section~\ref{sec:cpu-acc}, to compute high-quality paths and slices for tensor networks. Additionally, there is the possibility of reusing common calculations, as described in Ref.~\cite{Vincent2022jetfastquantum}, which avoids adding duplicate tasks during the slicing, therefore leading to a task graph with shared nodes. 

\subsubsection{Tensor network contraction}
A possible approach, presented for Jet~\cite{Vincent2022jetfastquantum} is to use the transpose-transpose-matrix-multiply method. The basic building block of the task-dependency graph is the pairwise contraction of two tensors. This can be decomposed into two independent (partial) tensor transposes and a single matrix multiplication~\cite{tensor-transpose-multiply}. While for the CPU they use the qFlex~\cite{qflex1, villalonga2020, Arute2019} transpose method, for the GPU they use cuTENSOR~\cite{bayraktar2023cuquantum}, used also in the work of Shah et al.~\cite{shah2023}. 

Another approach to tensor contraction is the bucket contraction algorithm by Dechter~\cite{bucket-elimination-alg}, as described in QTensor~\cite{lykov2021} and used also by Shah et al.~\cite{shah2023}, where an ordered list of tensor buckets (collection of tensors) is created to contract the tensor network. 
Each bucket corresponds to a tensor index, the bucket index. Buckets are then contracted one by one. The contraction of a bucket is performed by summing over the bucket index, and the resulting tensor is then appended to the appropriate bucket. The number of unique indices in aggregate indices of all bucket tensors is called a bucket width. Memory and computational resources of a bucket contraction scale exponentially with the associated bucket width. It is possible to improve this method by ordering the buckets first and then finding the indices that can be merged before performing the contraction. This allows for a smaller output size and larger arithmetic intensity as presented by Lykov et al.~\cite{lykov2021}. 
An issue with the bucket elimination algorithm is that tensors can grow too large to fit in memory, so a solution proposed by Shah et al.~\cite{shah2023} is to introduce data compression. 
Their work proposes a GPU-based lossy compression framework that can compress floating-point data stored in quantum circuit tensors with optimized speed while keeping the simulation result within a reasonable error bound after decompression. The compressed data can be decompressed when the tensors are needed during the computation. 

An additional approach is to use a partitioning method, which is optimal for each part of the circuit, depending on if it is sparse or dense, as mentioned already in the Schrödinger approach in the same work of Zhang et al. \cite{HyQuas}. OShareMem method~\cite{HyQuas}, an optimized version of the ShareMem method~\cite{sharemem}, deletes redundant computation, reduces data indexing overhead, and uses a new layout to access the shared memory faster. Another approach is the TransMM method~\cite{HyQuas}, which converts a set of special matrix-vector multiplications in quantum circuit simulation into GEMM to take advantage of highly optimized GEMM libraries and hardware-level GEMM compute units like Tensor Cores. 
The work of Zhang et al.~\cite{HyQuas}, as mentioned before, introduces an automatic selection method for the approach to be used, by using OShareMem for sparse parts of the circuit, and TransMM method for the dense parts of the circuit, allowing for further speedup.

\subsubsection{Data format optimisation}

As already mentioned for the CPU optimizations, using smaller precision for the data and calculations leads to improved performance. 
Although the works that try to accelerate the tensor-based acceleration on GPU do not try to propose any optimisation regarding the data format, they pick slightly different configurations. 
Most of the works use double (64-bit) floating-point precision~\cite{shah2023,willsch2022,HyQuas}, while other works, such as Jet~\cite{Vincent2022jetfastquantum}, use different precision such as single (32-bit) floating-point precision.

 \subsection{Other acceleration approaches}
 Some simulation tools allow for the acceleration of quantum circuit simulation on GPU by focusing more on the optimal usage of the available hardware. An example of this is cuQuantum~\cite{bayraktar2023cuquantum}, which provides composable primitives for GPU-accelerated quantum circuit simulation, including distributed computing on multiple GPUs. TensorLy-Quantum~\cite{patti2021tensorlyquantum} is a quantum library with direct support for tensor decomposition, regression, and algebra. It additionally provides built-in support for Multi-Basis Encoding for MaxCut problems~\cite{patti-multicut} and was used to develop Markov chain Monte Carlo-based variational quantum algorithms~\cite{patti-mql}.

% \section{FPGA for simulation acceleration}

\section{Acceleration using FPGA}
\label{sec:fpga-acc}
Although FPGAs are more limited in terms of resources and frequency when compared to CPUs and GPUs, they allow for more flexibility, leading to architectures better fitted to problem. While less scalable compared to the implementations with the previous hardware platforms, FPGAs excel when working with very limited resources, such as simulations running on an off-the-shelf personal computer.

It is worth noting that some of the reported work, especially early efforts, focused on emulating quantum hardware rather than serving as an acceleration platform for quantum computer simulation, where only a few core operations are offloaded to the FPGA. 

In the former case, the FPGA acts as a quantum computer itself, with which the host system must interface.
Some example of this emulation approach include the early works of Khalid et al.~\cite{khalid-2004} and Aminian et al.~ \cite{Aminian-2008}. This same approach has been followed more recently in terms of work of Zhang et al.~\cite{zhang2019fpga}. These approaches are interesting but can only be applied for problems requiring only a very limited number of qubits. 

More recent works focus on running only the most compute-heavy operations on the FPGA.
FPGA-accelerated simulation mainly focuses on the previously discussed Schrödinger approach. The two main steps of the acceleration are the pre-computation of the circuit matrix and the matrix computation.

\subsection{Pre-computation optimisation}
In the pre-computation step, various optimisations have been proposed. The work of Jungjarassub and Piromsopa~\cite{Jungjarassub2022} introduces an algorithm that does not directly multiply all the gates, but first checks for special cases. 
Depending on the combination of gate types and the transformation they apply, they can avoid executing certain multiplications in certain specific cases. 

A similar approach is used in the work of Hong et al.~\cite{hong2022quantumcircuitsimulator}, which checks for the matrix values before multiplication, and in the case where ones or zeros are present, avoids processing through the actual multiplication module. 
This work focuses on the computation of a $2 \times 2$ gate regardless of the circuit. It first groups multiple gates into a single one when possible, then simulates the resulting gates one by one. Although this approach does not prioritize parallel execution, it allows for the execution of larger circuits on hardware with limited resources.

\begin{table*}[htbp]
    \centering
    \caption{Hardware-aware optimization techniques.}
    \resizebox{\textwidth}{!}{%
 \begin{tabular}{ccc}
        \toprule
         \bf{Technique} & \bf{Improves} & \bf{Hardware support}\\
         \midrule
         Mixed-precision operations & compute \& data storage & CPU~\cite{precision_cpu}, GPU~\cite{precision_gpu}\\
         Arbitrary-precision operations & compute \& data storage & FPGA~\cite{precision_fpga}\\%\hline
         Lossy data compression & data storage & CPU (software), GPU (software), FPGA (compression engine)~\cite{lzma_engine}\\%\hline
         Thread-/Task-level parallelism & compute & CPU, GPU\\
         Vector (SIMD) instructions & compute & CPU with vector support~\cite{riscv_vpu}\\
         Matrix operations & compute & CPU with MMU~\cite{power10_mma}, GPU tensor cores~\cite{matrix_gpu}, FPGA Systolic Array~\cite{sa_fpga},\\ 
         & & AI engines~\cite{google_tpu}\\
         \bottomrule
    \end{tabular}
}
    \label{tab:hw_optimizations}
\end{table*}

\subsection{Computation optimisation}
Different approaches have been proposed to speed up the actual calculation. In the case of gates that generate phase rotation, it is possible to use a look-up table to collect sine and cosine values instead of calculating them directly~\cite{Jungjarassub2022}. Look-up tables are also evaluated in the work of Mahmud and  El-Araby~\cite{mahmud2018}, which replaces any complex calculations with a simpler array-indexed operation. This is further optimized by storing only the value relative to the actual input vectors which are necessary for the simulation. 

Another optimization to save memory operations is to check whether the state vector changes after a gate is applied before saving it back to memory. If no changes are detected, the memory-write can be avoided as is done in the work of Hong et al.~\cite{hong2022quantumcircuitsimulator}. 
 The work of Mahmud and  El-Araby~\cite{mahmud2018} evaluates different hardware architectures to identify the best-performing one for the Complex Multiply and ACcumulate (CMAC) operation, which is the core of simulation.
 They evaluate a single CMAC unit which processes all the operations (optimized for area but with low throughput), N-concurrent-CMAC units (optimized for throughput but requires a higher number of CMAC instances) and a dual-sequential-CMAC architecture (two CMAC instances connected sequentially, with computation and data write operations overlapped). Lastly, they also propose a kernel-based emulation model useful in case of a repeated set of core operations. The follow-up work by Mahmud et al.~\cite{mahmud2020} proposes an additional approach that improves the scalability of the emulation. Instead of using a look-up table, which sacrifices area for speed, or using dynamic generation (generation of the algorithm matrix elements at compile-time), which sacrifices speed for area, they propose a stream-based CMAC. They stream the algorithm matrix elements at run-time, meaning the operation's cost is typically the I/O channel latency between the control processor and the FPGA, which is negligible compared to the time required for processing the algorithm matrix.

\section{Summary and Future Directions}
\label{sec:summary}

In this work, we examined different approaches to speeding up and reducing the memory usage for the simulation of quantum computers. The most important results, reported in  Table~\ref{tab:summary-cpu} for the CPU-related works and in Table~\ref{tab:summary-gpu} for the GPU-related works, show how this is possible with multiple different approaches. 

Focusing on the hardware-aware optimizations for the simulation problem and challenges, we present in Table~\ref{tab:hw_optimizations} a summary of the different techniques, what they aim at improving, and which hardware support is best for that purpose. For each hardware support we also include a reference to a sample work showing a representative implementation of the technique in domains other than quantum computing.

The first group of optimizations aim at improving both the computation and data storage by changing the data and operations to use reduced precision. This has obvious benefits, but the challenge is to control the error introduced, since it is an optimization that loses information from the original 64-bit double floating-point precision. For these optimizations, FPGAs can offer the benefit to implement hardware that can operate on data of arbitrary precision (not necessarily conforming to standard sizes). Using reduced-precision data and operations is an optimization that is successfully being exploited in the Machine Learning (ML) domain currently by applying quantization of the data values~\cite{ml-quantization}.

The second group of optimizations are those focusing on reducing the data storage and are based on data compression techniques. In this case, since the goal is to reduce the data size considerably, lossless techniques are not enough and thus a lossy technique needs to be explored. This technique again comes with the price of reduced accuracy since there is a reduction of information after compressing the data. Several techniques have been applied in the past for scientific applications~\cite{sci-compression} and in order for the techniques to be applicable with reduced latency, it is important to have a hardware compression module~\cite{lzma_engine}. 

The third group of optimizations focus on lowering the execution time by leveraging parallel processing and/or using dedicated hardware units for some demanding operations. Thread and task parallelism is an effective technique to reduce the execution time by exploiting the existing parallel hardware resources in CPUs and GPUs. Single-Instruction-Multiple-Data (SIMD) instructions are used to exploit parallelism at a finer-grained scale, at the instruction level. When using SIMD instructions we are using multiple computational units to perform multiple operations in the same cycle. SIMD support is now common in most commodity processors, but the other operations that are very relevant to these simulations are matrix operations, which usually require dedicated matrix units for acceleration. Since matrix operations are also very relevant for AI, there has been a recent increase in products providing hardware support for these operations~\cite{dl-survey}. Namely, some CPUs and GPUs now have dedicated matrix units and/or neural acceleration units. Also, the Google TPU~\cite{google_tpu} used for both ML training and inference is basically a hardware accelerator for the matrix-matrix operations. The use of FPGAs in this case is also very relevant as they can be used to implement matrix-matrix operations for certain non-standard matrix dimensions. Also, as mentioned previously, FPGAs could be exploited to directly implement matrix-matrix operations for complex numbers~\cite{mahmud2018}.

Considering all of the above, we believe that in the future there is a need to invest in the development of more effective data compression techniques, a more flexible use of data precision. Since the market share for the simulation of quantum computers is much smaller than for AI applications, we do not expect an explosion of accelerators as has been observed in the recent years for the ML domain. As such, it is unlikely that we will see many hardware accelerators dedicated to the simulation of quantum computers. But the developments in the ML domain can be leveraged to deliver benefits to this domain too, since the critical operations are basically the same: matrix-matrix operations. As such, we expect in the future to see the use of ML accelerators for improving the performance of quantum computer simulations. Lastly, it is very interesting to already see some development using FPGAs and we expect the trend will be to see more and more FPGA-based dedicated hardware to  solve specific operations on specific data types in a very efficient way.

% \section{Conclusions}
\section{Conclusions}
\label{sec:conclusion}

Quantum computer simulators play an extremely important role in helping the development of new algorithms and hardware for the promising quantum computing paradigm.
While several approaches and simulators have been proposed, the characteristics of the problem make it extremely hard to scale to systems with a larger number of resources, required for solving more complex problems. 

The main challenges of quantum computer simulations are the large amount of data that must be stored in memory, with sufficient precision to produce results with acceptable accuracy, and the execution time, which grows exponentially with the increased number of qubits. 
In this work, we presented a review of existing tools and approaches for systems with CPUs, GPUs, and FPGAs, with a focus on how hardware-aware optimizations can help address the challenges.
Based on this study we showed the future directions for hardware-aware optimizations, including the use of accelerators designed for other domains but addressing similar problems and the development of FPGA-based hardware accelerators for quantum computer simulation.

% \section*{Acknowledgment}
\begin{acks}
We acknowledge support from the Swedish Foundation for Strategic Research (grant number FUS21-0063), the Horizon Europe programme HORIZON-CL4-2022-QUANTUM-01-SGA via the project 101113946 OpenSuperQPlus100, and from the Knut and Alice Wallenberg Foundation through the Wallenberg Centre for Quantum Technology (WACQT). AFK is also supported by the Swedish Research Council (grant number 2019-03696) and the Swedish Foundation for Strategic Research (grant number FFL21-0279). The work is also partially funded by the European High Performance Computing Joint Undertaking (JU) under Framework Partnership Agreement No 800928 and Specific Grant Agreement No 101036168 (EPI SGA2). The JU receives support from the European Union’s Horizon 2020 research and innovation programme and from Croatia, France, Germany, Greece, Italy, Netherlands, Portugal, Spain, Sweden, and Switzerland.    
\end{acks}

%\section*{References}
%\input{bibliography}
% \FloatBarrier
% \raggedright
% \bibliographystyle{unsrt}
\bibliographystyle{ACM-Reference-Format}
\bibliography{refs}

\end{document}